%% file: Main.tex
\documentclass[letterpaper]{article} % DO NOT CHANGE THIS
\usepackage{aaai2026}  % DO NOT CHANGE THIS
\usepackage{hhline}
\usepackage{xcolor}
\usepackage{colortbl}
\usepackage{amssymb}
\usepackage{multirow}
\usepackage{amsmath}
\usepackage{times}  % DO NOT CHANGE THIS
\usepackage{helvet}  % DO NOT CHANGE THIS
\usepackage{courier}  % DO NOT CHANGE THIS
\usepackage[hyphens]{url}  % DO NOT CHANGE THIS
\usepackage{graphicx} % DO NOT CHANGE THIS
\usepackage{natbib}  % DO NOT CHANGE THIS AND DO NOT ADD ANY OPTIONS TO IT
\usepackage{caption} % DO NOT CHANGE THIS AND DO NOT ADD ANY OPTIONS TO IT
\usepackage{algorithm}
\usepackage{algorithmic}

\usepackage{newfloat}
\usepackage{listings}
\DeclareCaptionStyle{ruled}{labelfont=normalfont,labelsep=colon,strut=off} % DO NOT CHANGE THIS
\floatstyle{ruled}
\newfloat{listing}{tb}{lst}{}
\floatname{listing}{Listing}
\title{Watermarking Visual Concepts for Diffusion Models}
\author{
    Liangqi Lei\textsuperscript{\rm 1},
    Keke Gai\textsuperscript{\rm 1},
    Jing Yu\textsuperscript{\rm 2},
    Liehuang Zhu\textsuperscript{\rm 1},
    Qi Wu\textsuperscript{\rm 3}
}
\affiliations{
    \textsuperscript{\rm 1}School of Cyberspace Science and Technology, Beijing Institute of Technology\\
    \textsuperscript{\rm 2}School of Information Engineering, Minzu University of China\\
    \textsuperscript{\rm 3}Australian Institute for Machine Learning, University of Adelaide\\
    3120245873@bit.edu.cn, gaikeke@bit.edu.cn, jing.yu@muc.edu.cn, liehuangz@bit.edu.cn, qi.wu01@adelaide.edu.au
}

\usepackage{bibentry}
\begin{document}

\maketitle

\begin{abstract}
The personalization techniques of diffusion models succeed in generating images with specific concepts. This ability also poses great threats to copyright protection and network security since malicious users can generate unauthorized content and disinformation relevant to a target concept.  Model watermarking is an effective solution to trace the malicious generated images and safeguard their copyright. However, existing model watermarking techniques merely achieve image-level tracing without concept traceability. When tracing infringing or harmful concepts, current approaches execute image concept detection  and model tracing sequentially, where performance is critically constrained by concept detection accuracy. In this paper, we propose a lightweight concept watermarking framework that efficiently binds target concepts to model watermarks, supporting simultaneous concept identification and model tracing via single-stage watermark verification. To further enhance the robustness of concept watermarking, we propose an adversarial perturbation injection method collaboratively embedded with watermarks during image generation, avoiding watermark removal by model purification attacks. 
Experimental results demonstrate that ConceptWM significantly outperforms state-of-the-art watermarking methods, improving detection accuracy by 6.3\%-19.3\% across diverse datasets including COCO and StableDiffusionDB. Additionally, ConceptWM possesses a critical capability absent in other watermarking methods: it sustains a 21.7\% FID/CLIP degradation under adversarial fine-tuning of Stable Diffusion models on WikiArt and CelebA-HQ, demonstrating its capability to mitigate model misuse.
\end{abstract}

 \begin{links}
     \link{Code}{https://anonymous.4open.science/r/Conceptwm}
     %\link{Datasets}{https://aaai.org/example/datasets}
     %\link{Extended version}{https://aaai.org/example/extended-version}
 \end{links}
\maketitle

\input{sec/1_intro}
\input{sec/2_related-work}
\input{sec/3_Preliminaries}
\input{sec/4_methodology}

\input{sec/5_experiment}
\input{sec/6_conclusion}

\bibliography{aaai2026}

% \newpage

\end{document}

%% file: sec/1_intro.tex
\section{Introduction}
\label{sec:intro}
%扩散模型近年来的研究推动了图像生成领域，使得通过简单提示词生成逼真的描述称为可能。
%扩散模型的个性化技术在生成特定主题驱动或风格驱动的概念描述方面取得了显著成效。借助强大的个性化工具，如DreamBooth和Lora，任何人，即使没有设计技能，也可以通过几张个人图像和简单的文本输入创建特定概念的图像。

Recent advance in diffusion models have significantly improved image generation ability, enabling the creation of high-quality images from personalized textual descriptions. 
The upgraded diffusion models achieve few-shot learning for new visual concept generation, such as unique visual themes, motifs, stylistic preferences, or celibrities, using only a few images with the target concepts \cite{shi2024instantbooth,gal2023encoder,gal2022image,kumari2023multi}.
However, these techniques pose significant risks for visual concept infringement and misuse. For example, a malevolent actor might leverage a few online trademark or digital artwork images to fine-tune off-the-shelf diffusion models and edit the content without permission \cite{zhu2024watermark,li2025towards}. Similarly, a few online celebrity images are adequate for malevolent  actors to generate misinformation with these celebrities \cite{wang2024simac}. It becomes an urgent issue to effectively trace AI-edited visual concept images for copyright protection and privacy preservation.

\begin{figure}[t]
\centering
\includegraphics[width=0.45\textwidth]{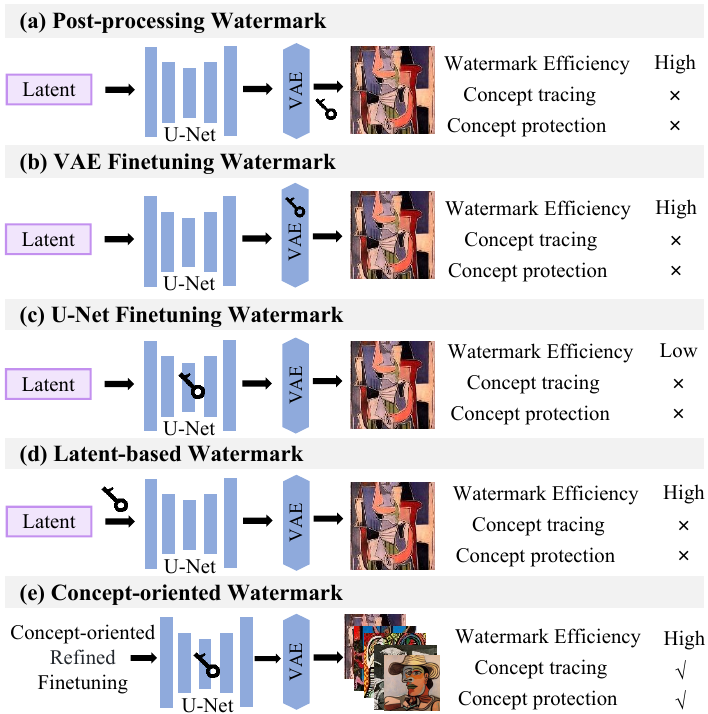} % 
\caption{Five main kinds of diffusion model watermarking methods: (a) Post-processing watermark. (b) VAE finetuning watermark. (c) U-Net finetuning watermark. (d) Latent-based watermark. (e) Concept-oriented watermark. }
% \yj{Image sample should better contain obvious `concept' instead of common object (cow).}
\label{fig1}
\end{figure}

Model watermarking is a representative way to trace the source model of generated images and thus verify the accountability of malicious content. The typical solutions of model watermarking fine-tune the pre-trained diffusion models to inject watermarks and extrect the watermark information from any generated images by a watermark decoder. Compared with post processing solutions by adding watermarks after image generation (Fig. \ref{fig1}(a)), model watermarking is more secure to avoid watermark evasion \cite{zhao2023generative}. Despite the great progress in diverse model watermarking schemes (Fig. \ref{fig1}(b)(c)(d)), existing schemes indiscriminately inject watermarks into all the generated images agnostic to visual concepts, which has two obvious limitations for visual concept tracing and protection. On the one hand, tracing concept-level malicious image becomes indirect that requires a two-stage pipeline: initial detection of target visual concepts in images, followed by watermark-based tracing for concept images. Thus, the tracing effectiveness is critically constrained by concept detection accuracy. One the other hand, recent studies \cite{cui2025ft} reveal that existing personalization methods enable malicious actors to remove watermarks by fine-tuning diffusion models with a few watermarked samples, while retaining concept generation capabilities. Therefore, current model watermarking methods with only image tracing capability are still unable to prevent misue of watermarked images.

In this paper, we propose an approach that aims to trace and protect generated images with target visual concepts by diffusion model. The core idea of our approach is \textit{Concept Watermarking}, that is, during image generation, we embed both traceable watermarks and fine-tuning-resistent perturbations  into concept-specific generated images. This paradigm enables people to technically govern AI-powered malicous content and proactively defend subsequent misuse of generated content in a finer manner. To this end, three new challenges need to be addressed: First, for a target concept (\textit{e.g.}, `a specific kind A of cats'), concept watermarking requires to establish tripartite correlation among a concept watermark (\textit{e.g.}, a bit sequence `01100'), visual concepts (diverse appreances of `A cats') in the generated images, and a textual concept (a psuedo word of `cat A') in the prompts, which is quite different from existing watermarking methods that solely establish bipartite correlation  between a watermark and model outputs; Second,  concept watermarking has adversarial capability that the generated concept images are unable to be used for model fine-tuning, which has no influence on concept-irrelevant generated images; Third, concept watermarking is robust to diverse real-world attacks including image processing, geometric transformations, and model regeneration attacks.

To address these challenges, we propose a lightweight concept watermarking framework, named ConceptWM, which contains two main componets, \textit{i.e.}, efficient concept watermark training and imperceptible adversarial perturbation injection. 
%To achieve the concept watermark embedding, we develop a Efficient Concept Watermark Finetuning (ECWF) scheme, which learns watermark patterns in diffusion models with limited data and training resources. 
To embed a watermark that is both fidelity-preserving and conducive to learning by U-Net, we propose a fidelity-preserving latent watermark module that incorporates the watermark into the latent space and introduce a fusion layer after the message encoder and latent concatenation to mitigate rigid watermark pattern artifacts. 
To achieve concept-level image tracing, we propose an efficient concept watermark training strategy that alternately optimizes DreamBooth fine-tuning and watermark embedding training to achieve joint concept-watermark learning. Our method learns watermark embeddings efficiently, requiring only a few reference images and low training budgets.
% Additionally, we develop a Fidelity-preserving Latent Watermark (FLW) module  that incorporates the watermark into the latent space and employs a distortion layer in order to enhance robustness against image processing attacks.
To prevent the misuse of watermarked images, we propose an imperceptible adversarial perturbation injection  module to refine both referenced model parameters and adversarial perturbations by proposed optimization strategy, which balances adversarial robustness, watermark fidelity, and image quality, while mitigating the attack risk by personalized fine-tuning.

Our contributions are as follows:
(1) We propose a lightweight concept watermarking framework that imperceptibly embeds  watermarks into the generated concept-specific images by diffusion models. Concept watermarking first tackles the challenge of establishing tripartite correlation across three modalities, \textit{e.g.,} watermarks, visual concept and textual concept, which makes it possible for concept-level generated image tracing in one-stop manner. 
%The proposed framework highlights the importance of refined watermark training and fidelity-preserving perturbations in real-world applications of diffusion model watermarking.
%We consider that a refined watermark training and fidelity preserving perturbation are two critical factors in the context of concept watermarking, which offers a fresh perspective on how diffusion model watermarking can be applied in real-world scenarios.
(2) We propose a new adversarial mechanism  that endows concept watermarks with proactive denfence capabilities. 
This mechanism enables the adversarial perturbation injection without interference to image quality and watermark fidelity, while making the generated concept images resistent to malicious use for model fine-tuning.
(3) ConceptWM achieves 6.3\% higher robustness against image processing than 4 baselines while maintaining a critical property: it forces 21.7\% FID/CLIP degradation in fine-tuned Stable Diffusion models - a capability beyond current methods.
%Experimental results demonstrate that ConceptWM maintains better robustness in watermark verification compared with existing 4 baselines over 6.3\% image processing attacks and 

%Experimental results demonstrate that ConceptWM effectively integrates adversarial perturbations into watermarks and maintains better robustness in watermark verification compared with existing 4 baselines over 6.3\% image processing attacks. 
\vspace{-0.1cm}
% Our contributions are as follows: (1) We have highlighted the vulnerabilities and limitations of current watermarking methods for diffusion models in protecting specific concepts, and proposed a Diffusion model watermark for protecting specific concepts in diffusion models. (2) We propose a modulation method that ensures a balance between watermark robustness and fidelity. (3) Extensive experiments and ablation studies demonstrate that our watermark satisfies all the previously stated requirements, as well as the effectiveness of the proposed designs.

% \yj{(1) concept watermark framework. (2) Technical contribution for a balance of robustness and  fidelity. (3) Good performance. Each contribution should give novel mechanism explanation or solid experimental results from various apects.}

\begin{figure*}[t]
\centering
\includegraphics[width=2.1\columnwidth]{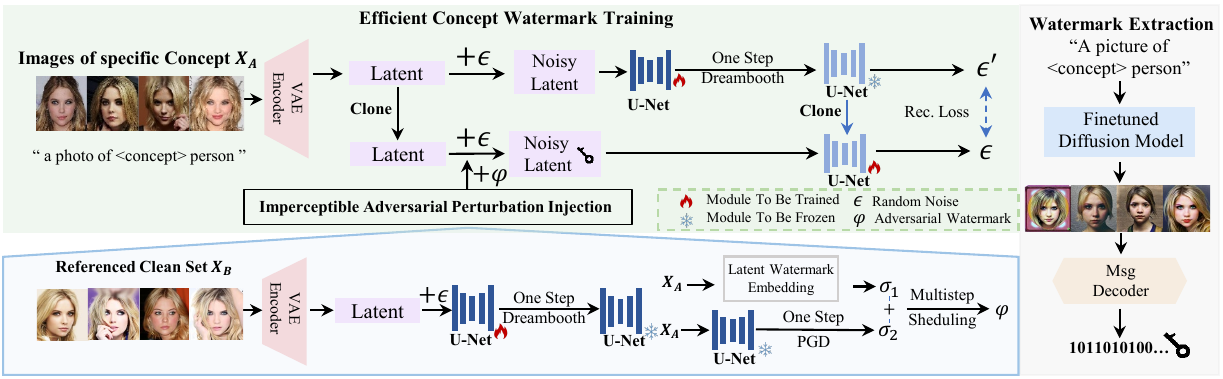} % 
\caption{Framework of ConceptWM. 
The ECWT module employs a two-step training process to learn specific concepts and watermark patterns. 
The IAPI module utilizes a surrogate model fully fine-tuned on a referenced clean image to guide PGD optimization, iteratively generating adversarial watermarks through multiple rounds of watermark and noise integration. }
\label{fig2}
\end{figure*}

%% file: sec/2_related-work.tex
\section{Related Work}

\noindent\textbf{Personalization Techniques for Diffusion Models.} 
Recent explorations mainly focus on injecting specific concepts into the model through fine-tunning tools, \textit{e.g.,} Textual Inversion \cite{gal2022image}, DreamBooth \cite{ruiz2023dreambooth} and LoRA (Low-Rank Adaptation) \cite{hu2021lora}.
%, given the substaintial computational requirements of training Stable Diffusion from scratch. 
Specifically, Textual Inversion focuses on training text embedding so that concepts are allowed to be injected into the text encoder without altering weights of the U-Net. 
% Dissimilar to conventioanl text-to-image fine-tuning, 
DreamBooth fine-tunes the entire U-Net portion of the Stable Diffusion model and employs prior loss during training to prevent overfitting. 
LoRA is designed to achieve lightweigth efficient fine-tuning by training incremental weights in the attention layers of the U-Net. 
Our work focuses on watermarking specific concepts generated through DreamBooth fine-tuning.
%Given the substantial computational requirements of training Stable Diffusion from scratch, many methods aim to inject specific concepts into the model through fine-tuning.
%Currently used fine-tuning techniques include Textual Inversion \cite{gal2022image}, DreamBooth \cite{ruiz2023dreambooth} and LoRA (Low-Rank Adaptation)\cite{hu2021lora}. 
%Textual Inversion focuses solely on training text embeddings during fine-tuning, allowing concepts to be injected into the text encoder without altering the weights of the U-Net. 
%DreamBooth, on the other hand, fine-tunes the entire U-Net portion of the Stable Diffusion model. 
%Unlike conventional text-to-image fine-tuning, DreamBooth employs prior loss during training to prevent overfitting. 
%LoRA achieves quick and lightweight fine-tuning of Stable Diffusion by training incremental weights in the attention layers of the U-Net. 

\noindent\textbf{Model Watermarking for Latent Diffusion Models.} 
The embedding characteristics of different watermarking methods for diffusion models are illustrated in Figure \ref{fig1}, 
%which is mainly divided into Post-processing Watermark, U-Net Finetuning Watermark, VAE-Finetuning Watermark, and Latent-based Watermark. 
These watermarking methods aim to protect the copyright of images or generative models, so they often embed watermarks into all images.
WDM \cite{zhao2023recipe} is designed to train an autoencoder to stamp a watermark on all training data before re-training the generator from scratch.
Aqualora \cite{feng2024aqualora} is a lora-based method that watermarks all images.
VAE-based methods \cite{xiong2023flexible,kim2024wouaf,fernandez2023stable} fine-tune VAE-Decoder to ensure all generated images containing the watermark.
However, these methods fail to protect specific concepts. 
% A large volume of images are generally required for the model to learn the watermark embedding pattern, but concept-oriented images for finetuning are limited to only a few. 
Our Efficient Concept Watermark Training embeds watermark into specific concept with a few images of specific concept.

% WDM \cite{zhao2023recipe} trains an autoencoder to stamp a watermark on all training data before re-training the generator from scratch.
% Aqualora \cite{feng2024aqualora} propose a watermark Lora method to watermark all the images. 
% VAE-based methods \cite{xiong2023flexible,kim2024wouaf,fernandez2023stable} fine-tune VAE-Decoder to ensure that all generated images contain the watermark.
% However, the aforementioned watermarking schemes apply watermarks to all generated images, leaving a gap in the protection of specific styles. 
% Meanwhile, A large number of images is often required for the model to learn the watermark embedding pattern, but concept-oriented images for finetuning are often limited.

\noindent\textbf{Protective Perturbation against Stable Diffusion.}
Recent studies \cite{wu2023towards,ye2023duaw,zhao2023unlearnable,zheng2023understanding} mostly have addressed disrupting the fine-tuning process by adding imperceptible protective noise to images, in order to prevent personal images from potential infringement during the fine-tuning of Stable Diffusion. 
AdvDM employs adversarial noise to safeguard personal images on Stable Diffusion by maximizing the Mean Squared Error loss during optimization \cite{liang2023adversarial}. 
Anti-DreamBooth \cite{van2023anti} offers a bi-level min-max optimization for generating protective perturbations, through integrating the DreamBooth fine-tuning process of Stable Diffusion into its framework. 
%Other studies \cite{wu2023towards,ye2023duaw,zhao2023unlearnable,zheng2023understanding} have explored similar adversarial perturbation techniques to generate protective noise for images. 
Our Imperceptible Adversarial Perturbation Injection effectively endows the watermark with the aforementioned adversarial properties.

%% file: sec/4_methodology.tex
\section{Methodology}

The main objective of our scheme is to protect images of a specific concept that are generated by the diffusion model, by embedding an adversarial watermark pattern into the U-Net component of the diffusion model. 
Fig. \ref{fig2} illustrates an overview of our approach. 
To embed a watermark that is both fidelity-preserving and conducive to learning by U-Net, we develop a Fidelity-preserving Latent Watermarking (FLW) module, which adds a combination layer for the message encoder and latent concatenation, so that watermarks can be generated based on image characteristics rather than a fixed watermark signal.
%allowing watermarks to be generated based on image characteristics rather than a fixed watermark signal. 
To protect the watermarked concept-specific images from misusing, we develop an Imperceptible Adversarial Perturbation Injection module to integrate adversarial propeties into the watermark. 
To enable U-Net to efficiently learn watermark patterns during concept learning with limited samples, we present an Efficient Concept Watermark Finetuning module that alternates optimization of model parameters to simultaneously achieve adversarial watermark embedding and concept learning.

\subsection{Fidelity-preserving Latent Watermark}
\label{latentwm}

% \begin{figure}[t]
% \centering
% \includegraphics[width=0.47\textwidth]{img/wm2v2.pdf} % 
% \caption{The comparison of two variants of Latent watermark namely Overlay-based Latent Watermarking and Fusion-based Latent Watermarking.}
% \label{comp}
% \end{figure}

Existing image watermarking techniques frequently arise substantial discrepancies between the watermark embedding pattern and the latent space. 
This mismatch issue may cause the disruption or even complete loss of watermark information when the image is encoded into the latent space by a VAE encoder or regenerated through diffusion, as has been discussed by recent watermarking attacks \cite{an2024benchmarking,balle2018variational,cheng2020learned}. Thus, we aim to embed a robust watermark within the image that is both resilient and conducive to efficient learning for U-Net. To be specific, we have developed two variants of latent watermarking, which are Overlay-based Latent Watermarking (OLW) and Fidelity-preserving Latent Watermarking (FLW). Both variants involve the latent domain perturbations. The watermarked images are processed through the message decoder to evaluate the discrepancy between the decoded watermark information and the original content of the watermarked image. 
The message loss is the Binary Cross Entropy (BCE) between the message $m$ and $\sigma(m')$. Given the message length N, the message loss $\mathcal{L}_m$ is defined in Eq. (\ref{bce}).
%The message loss is the Binary Cross Entropy (BCE) between the message $m$ and the sigmoid $\sigma(m')$. 
\begin{equation}
\begin{aligned}
    \mathcal{L}_m
    =-\sum_{k=1}^{N}m_k\log\sigma(m_k')+(1-m_k)\log(1-\sigma(m_k'))
\end{aligned}
\label{bce}
\end{equation}

\noindent\textbf{Overlay-based Latent Watermarking.} OLW essentially overlays a fixed watermark signal that can be removed through subtraction. 
We observe that OLW-watermarked images may face the issue of the watermark integrity, since noticeable artifacts appear in local regions of the images, making the images distinguishable to the huaman eye. 

%the images appear noticeable artifacts in local regions, making the images distinguishable to human eyes. 

\noindent\textbf{Fidelity-preserving Latent Watermarking.} 
To address issues of OLW, we have developed Fidelity-preserving Latent Watermarking (FLW) that adds a combination layer after the message encoder and latent concatenation. 
To ensure visual consistency, we calculate LPIPS loss $\mathcal{L}_{pips}$ \cite{zhang2018unreasonable} and utilize a Peak Pixel Difference (PPD) Loss $\mathcal{L}_{ppd}$ which selects the most prominent parts of the image and calculates channel loss to reduce excessive modification of the images.
Our training objective ($\mathcal{L}_{total}$) can be expressed in Eq. (\ref{eq:ltotoal}), where $\lambda$ and $\mu$ represent the coefficients.
\begin{equation}\label{eq:ltotoal}
\mathcal{L}_{total}=\mathcal{L}_m+\lambda\mathcal{L}_{pips}+\mu\mathcal{L}_{ppd}
\end{equation}
The jointly optimized message encoder and decoder, working in concert with the Adcersarial Perturbation Modulation module, will empower the U-Net to learn robust watermarking patterns. We provide the training structure comparison of OLW and FLW in the Appendix.

\subsection{Imperceptible Adversarial Perturbation Injection}
\label{Modulation}
To protect the generated concept-specific images from removing through personalized fine-tuning, we integrate adversarial properties into the watermark. 
By maximizing the image generation loss with adversarial noise added to the watermark and minimizing the concept learning loss through Dreambooth fine-tuning with reference images, we ensure that the watermark does not interfere with concept learning while preventing the misuse of watermarked image.
% preventing ``jailbreaking'' through personalized fine-tuning.

\noindent\textbf{Image Generation Loss.} 
To enable the conditional denoising model $\epsilon_\theta(x,t,c)$ to match the data distribution of the target images, $\mathcal{L}_{cond}$ is a loss function that ensures the conditional denoising model generates images conditioned on a given text condition ($c$). 
Given the input image $x_0$, the $\mathcal{L}_{cond}$ is as follows in Eq. (\ref{diff1}):
\begin{equation}
\resizebox{0.9\hsize}{!}{
   $ \mathcal{L}_{cond}(\theta,x_0)=\mathbb{E}_{x_0,t,c,\epsilon\in\mathcal{N}(0,1)}\|\epsilon-\epsilon_\theta(x_{t+1},t,c)\|_2^2$
    }
\label{diff1}
\end{equation}
where each variable $x_t$ is constructed through noise injection at a corresponding timestep $t$ within the set $\{1,2,\cdots, T\}$.

\noindent\textbf{Concept Learning Loss.} 
This ensures conditional denoising model $\epsilon_\theta(x,t,c)$ to learn the concept of specific images. 
Dreambooth \cite{ruiz2023dreambooth} uses a prior prompt such as ``a photo of [class]" and learns the concept with a prompt such as ``a photo of $\langle$concept$\rangle$ [class]". DreamBooth employs the concept loss ($\mathcal{L}_{ft}(\theta,x_0)$) is expressed in Eq. (\ref{eqa2}).
\begin{equation}
\resizebox{1\hsize}{!}{
$\mathcal{L}_{ft}(\theta,x_0)=\mathbb{E}_{x_0,t}\|\epsilon-\epsilon_\theta(x_{t+1},t,c)\|_2^2 +\lambda\|\epsilon'-\epsilon_\theta(x_{t+1}',t,c_{cls})\|_2^2$
}
\label{eqa2}
\end{equation}
where $\epsilon$ and $\epsilon'$ are both sampled from the standard normal distribution $\mathcal{N}(0,\mathbf{I})$,
The noisy variable $x_{t + 1}'$ is generated from the original stable diffusion model based on the prior prompt $c_{\text{cls}}$.
$\lambda$ emphasizes the importance of the prior term in the generation process.
The term $\lambda$ emphasizes the importance of the prior term in the generation process.

% The adversarial requirement for watermarking is to disrupt the learning process of personalized finetuning. Since the personalized finetuning model tends to overfit on adversarial images, this can be exploited to make it perform worse when reconstructing clean images. 
% \begin{equation}
% \begin{aligned}
% \sigma^{*(i)} & =\underset{\delta^{(i)}}{\arg \max } \mathcal{L}_{\text {cond }}(\theta^{*}, x^{(i)}), \forall i \in\{1, . ., N_{ft}\}, \\
% \text { s.t. } & \theta^{*}=\underset{\theta}{\arg \min } \sum_{i=1}^{N_{ft}} \mathcal{L}_{ft}(\theta, x^{(i)}+\sigma^{(i)}), \\
% \text { and } & \|\sigma^{(i)}\|_{j} \leq \eta \quad \forall i \in\{1, . ., N_{ft}\},
% \end{aligned}
% \end{equation}
% where $x^{(i)}$ denotes the clean samples. The adversarial noise \(\sigma^{*(i)}\) is within the constraint range of $\eta$.
% % Meanwhile, we need to ensure that the image quality and watermark loss are as minimal as possible. 

\noindent\textbf{Adversarial Perturbation Injection.} 
The goal of this modulation is to maximize $L_{cond}$ in Eq. (\ref{diff1}) to hinder personalized finetuning and minimize $L_{ft}$ in Eq. (\ref{eqa2}) to learn the concept. 
Our solution is to use watermark-free images $\mathcal{X}_A$ for one-step DreamBooth training to reduce the DreamBooth loss $\mathcal{L}_{\text{ft}}$, while simultaneously applying Projected Gradient Descent (PGD) \cite{madry2017towards} to enhance the adversarial loss $L_{cond}$. 
The model used for PGD is trained with clean referenced images $\mathcal{X}_B$ in a one-step dreambooth training process, rather than the images used for concept learning. 
The purpose of this approach is to allow the model to learn adversarial perturbations without affecting the learning of the concept. The specific procedure is defined by Eq.s (\ref{eqa_adv1}) and (\ref{eqa_adv2}), where $\sigma_1$ denotes the residual of the watermarked image and original image. $x^i$ denotes the clean samples. 
\begin{equation}
\theta^*=\arg\min_{\theta}\sum_{x\in\mathcal{X}_A}\mathcal{L}_{ft}(\theta,x^{i})
\label{eqa_adv1}
% &\theta^{*} &&\leftarrow \theta.\text{clone}()\\
\end{equation}
\begin{equation}
\varphi^{(i)}=\arg\max_{\sigma_2^{(i)}}\mathcal{L}_{cond}(\theta^{*},x^{(i)}+\sigma_1+\sigma_2^{(i)}) 
\label{eqa_adv2}
\end{equation}
The adversarial noise \(\sigma^{*(i)}\) is within the range of $\eta$. The Imperceptible Adversarial Perturbation Modulation outputs an adversarial watermark $\varphi$ as the parameter for Concept-oriented Watermark Training. The pseudo-code of Adversarial Perturbation Modulation is presented in the Appendix.

% The pseudo-code of Adversarial Watermark Modulation is presented in \ref{algorithm: cwf}.
% Considering that the adversarial perturbation will affect the watermark, we also incorporate the decoded message loss as a constraint to optimize the perturbation.
% \begin{equation}
% \mathcal{L}_{msg}=BCE(decoder(x+\sigma_1+\sigma_2^{(i)}),m_{target})
% \end{equation}

\subsection{Efficient Concept Watermark Training}
\label{fine-tune}
% At this stage, the watermark is embedded into specific concepts and then injected into the U-Net. The main challenges are as follows: The fine-tuned images for specific concepts are often limited, making it difficult for the model to learn the specific watermark patterns; Due to the introduction of adversarial perturbations, directly using the DreamBooth training paradigm will be affected by protective perturbations, preventing the model from learning the concept.

% To enable U-Net to efficiently learn watermark patterns during concept learning with limited samples, 
We aim to integrate adversarial watermarking into concept learning in this module.
The primary challenges are as follows: The availability of fine-tuned images for specific concepts is often limited, hindering the model's ability to learn the distinct watermark patterns. 
The introduction of adversarial perturbations can interfere with the DreamBooth training process, as these protective perturbations prevent the model from effectively learning the concept.
Our core idea is to decouple concept learning from adversarial watermark learning. 
In the DreamBooth stage, the U-Net learns the specific concept. 
%In the adversarial watermark learning stage, the model is fed with both watermarked and non-watermarked images, and the difference in the predictions is used as a loss to guide the generative model in learning the watermark embedding pattern.
In the adversarial watermark learning stage, the model is fed with both watermarked and non-watermarked images, and by using the difference in the predictions as a loss, it guides the generative model in learning the watermark embedding pattern.

% The limited number of specific concept samples and adversarial perturbation often result in failures when directly adding an adversarial watermark into images and deploying DreamBooth. This can prevent the watermark embedding pattern from being learned effectively and lead to quality degradation. Our core idea is to decouple concept learning from adversarial watermark learning. 
% In the DreamBooth stage, we learn the specific concept, and by adding adversarial watermark to the same image and also using the image without the adversarial noise, the model learns to differentiate between them. 
The training objective of this stage is defined by Eq. (\ref{eq:l}), where $\varphi$ represents the adversarial perturbation containing the watermark generated through IAPI, $\epsilon_{\theta}^{*}$ denotes the fixed model, and $\epsilon_\theta$ refers to the trained model. 
\begin{equation}\label{eq:l}    
\begin{aligned}\mathcal{L}_{\mathrm{cwft}}(\theta,x_t)&=\mathbb{E}_{t,c,x}\Big[\|\boldsymbol{\epsilon}_\theta(x_t+\varphi,t,c)-\boldsymbol{\epsilon}_{\theta}^{*}(x_t,t,c\big)\|^2\Big]\end{aligned}
\end{equation}

The pseudo-code of Concept-oriented Watermarking Training is presented in the Appendix. 
% $\varphi$ represents the adversarial perturbation containing the watermark generated through Adversarial Watermark Modulation, $\epsilon_{\theta}^{*}$ denotes the fixed model, and $\epsilon_\theta$ refers to the trained model. 

\noindent \textbf{Prompt Adaption.} 
Our goal is to preserve the concepts embedded in the images used for DreamBooth fine-tuning. During DreamBooth training, the prompt is limited to ``a photo of $\langle$concept$\rangle$ [class].'' However, 
in practical applications, users might apply different prompts incorporating the concept term to generate images. This can lead to significant distribution differences between the generated images and the watermark-embedded training images, potentially causing performance degradation. 
To address this issue, 
%we propose a simple yet effective solution. 
%after the Dreambooth fine-tuning is completed, 
we simulate real-world scenarios by generating images that combine the $\langle$concept$\rangle$ with different prompts and then perform Concept Watermark Training in the same manner.
% %with different prompts combined with the  $\langle$concept$\rangle$ and then perform Concept-oriented Watermark Fine-tuning in the same manner. 
% This strategy effectively enhances performance. 
% The effectiveness can be found in ablation Section \ref{ablation}.

% The pseudo-code of Concept-oriented Watermarking Fine-tuning is presented in \ref{algorithm: cwf}. 

%% file: sec/5_experiment.tex
\begin{figure*}[t]
\centering
\includegraphics[width=1.85\columnwidth]{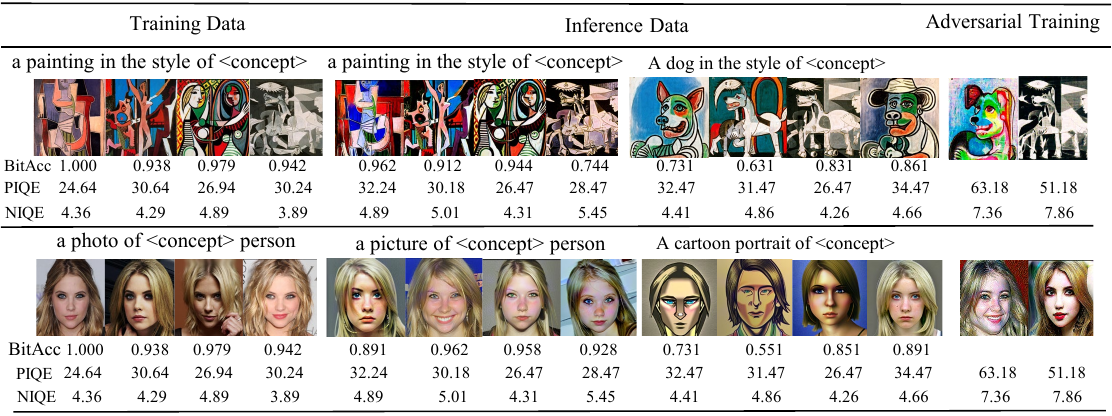} % 
\caption{Demonstration of the generated images. Below each image, we display the corresponding Bit Accuracy, along with the image quality metrics PIQE and NIQE.}
\label{fig3}
\end{figure*}
\input{tab/table1}
\section{Experiment}

 \noindent\textbf{Datasets.} To evaluate the effectiveness of the proposed methods, for subject-driven generation, we select the face dataset CelebA-HQ \cite{zhu2022celebv}.
 For style-driven generation, we select 12 artists with distinct styles from WikiArt \footnote{https://www.wikiart.org}, with each artist contributing 10 artworks that are consistent in terms of both artistic style and time period. Each image is resized and center-cropped to a resolution of 512 × 512. In each experiment, a total of 5–8 carefully selected images were utilized for DreamBooth fine-tuning.

\noindent\textbf{Training Configurations.}
By default, we use the Stable Diffusion v2.1 as the pretrained generator. Unless stated otherwise, the training instance prompt for subject-driven generation is ``a photo of $\langle$concept$\rangle$ person" and the prior prompt is ``a photo of person". The training instance prompt for style-driven generation is ``a painting in the style of $\langle$concept$\rangle$" and the prior prompt is ``a painting". For PGD, we use $\alpha$ = 1e-5 and the default noise budget $\eta$ = 1e-3. The learning rate of the dreamBooth is set 5e-6. 

\input{tab/table2}

\noindent\textbf{Evaluation Metrics.}
We evaluate the detection performance using the true positive rate (TPR) at a false positive rate (FPR) of 10e-5. For traceability performance, we assess bit accuracy. To measure the quality of image generation, we compute the Peak Signal-to-Noise Ratio (PSNR) \cite{hore2010image} and Structural Similarity Index (SSIM) \cite{wang2004image}; and the Natural Image Quality Evaluator (NIQE) \cite{mittal2012making} and Perceptual Image Quality Evaluator (PIQE) \cite{venkatanath2015blind} for evaluating overall image quality. 
% For the semantic performance metrics, we adopt the CLIP score \cite{Radford2021LearningTV} and the Fréchet Inception Distance (FID) \cite{heusel2017gans}, calculated on the COCO2017 validation set.

\noindent\textbf{Watermarking baselines.} We select four typical baselines: two official watermark of Stable Diffusion \cite{rombach2022high} called DwtDctSvd \cite{cox2007digital}, and RivaGAN \cite{zhang2019robust},VAE-based method Stable Signature \cite{fernandez2023stable} and Latent-based method Tree-ring \cite{wen2024tree}. DwtDctSvd, RivaGAN, and Stable Signature all use 48-bit watermarks, while Tree-ring uses P-value as the evaluation method.

\subsection{Watermark Performance Comparison}
We evaluate our method against 4 typical baselines on two tasks: 
(1) Detection: All methods are tested as single-bit watermarks with a unified watermark, with the False Positive Rate (FPR) set to $10^{-5}$, and we assess the True Positive Rate (TPR) on 1,000 watermarked images. 
(2) Traceability: Except for the single-bit watermark Tree-Ring, all other methods are tested as multi-bit watermarks. 
Specifically, we simulate a scenario with 1,000 model users, each requiring a distinct watermark for model tracing. 
Each user generates 10 images, creating a dataset of 10,000 watermarked images. 
During testing, we calculate the Bit Accuracy by comparing the watermark of each user with the watermarked images generated by the user. 
%The user with the highest Bit Accuracy is considered the traced user. 

The comparison results, shown in Table \ref{tab1}, demonstrate that our watermarking method achieves superior robustness, significantly outperforming the baselines in both tasks. 
In terms of Bit Accuracy, our method outperforms the best baseline by around 6.3\%. OLW exhibits strong performance in robustness experiments. However, it suffers from relatively unsatisfactory image quality (FID 24.06,CLIP 0.357). FLW demonstrates a relatively balanced result, maintaining good image quality while ensuring strong robustness. We found that there is a trade-off between image quality, robustness, and detection accuracy in watermark embedding. Due to extensive adversarial training on the watermark model, FLW achieves a bit accuracy of 95.68 in a non-attack environment, which is relatively lower compared to other watermarking schemes. To further improve bit accuracy in a non-attack setting, it would require considering either lower image quality or reduced robustness.
% The robustness against each type of attack can be seen in Table \ref{tab2}.
% This improvement derives from the widespread diffusion of the watermark throughout the latent space, creating a strong binding between the watermark and the image semantics. 

\subsection{Watermarked Diffusion Model Fidelity}
For fidelity evaluation metrics, we use the Fréchet Inception Distance (FID) \cite{heusel2017gans} for assessing image diversity and semantic relevance, which is calculated on 5000 images from the COCO 2017 validation set \cite{lin2014microsoft}. % to assess image quality. 
Additionally, we use CLIP score \cite{Radford2021LearningTV}, a method for measuring image similarity, which provides results more aligned with human judgment. Table \ref{tab1} presents a comparison of our method with other baseline methods. 
\input{tab/table3}

\noindent\textbf{Main Results}.
% The performance of this method is relatively lower compared to other methods. 
The performance of our method outperforms the state-of-the-art methods, with minimal performance difference compared to the original model (FID difference of 0.21, CLIP difference of 0.02). This method is a localized 
fine-tuning of the model and does not significantly impact the model's overall performance. Due to the smaller dataset and the more refined watermarking method, our method has a smaller impact on model generation quality compared to other methods. It can be observed that, among the two variants of latent watermarking introduced, OLW and FLW, the image quality produced by OLW, with an FID of 24.06 and semantic quality (CLIP) of 0.357, is comparatively lower. In contrast, the image quality produced by FLW, has an FID of 24.58 and semantic quality (CLIP) of 0.365. This disparity arises because the watermark in FLW is contextually related to the image content, whereas the watermark in OLW is a fixed, image-independent watermark, which leads to degradation in image quality.

\subsection{Watermark Robustness against Attack}

\noindent\textbf{Image Processing.} We select eight representative types of image-level noise to evaluate the robustness of image processing: (1) Brightness, (2) Contrast, (3) Blur, (4) Gaussian Noise, (5) JPEG Compression, (6) Crop \& Scale, (7) VAE Compression \cite{cheng2020learned}, and (8) Diffusion-based Reconstructive Attack \cite{zhao2023generative}. These cover common image processing techniques, geometric transformations, and regeneration attacks encountered in real-world scenarios. The main results are shown in Table \ref{tab2}.

\input{tab/table4}

\noindent\textbf{Main Results.} 
For each image processing attack, we report the average bit accuracy/ TPR@1e-5FPR in Table \ref{tab2}, which shows ConceptWM is indeed robust  across all the transformations with the bit accuracy consistently above 0.9. 
ConceptWM remarkably outperforms existing multi-bit watermarks on the average performance with more than 6.3\% boost compared with the state-of-the-art (SoTA) results. Since our method embeds the watermark into the entire latent space and trains the distortion layer, it has a clear advantage over contrastive watermarking schemes when facing various image processing attacks. 
% Please refer to the Supplementary Materials for more results.

\input{tab/table5}

\subsection{Concept Protection Evaluation}

\noindent\textbf{Model Mismatching.}
We provide an example
of transferring adversarial watermark trained on SD v1.4 and SD v2.1 to defend
DreamBooth models trained from v2.1 and v1.4 in Table \ref{tab4}. The results show that our method still works effectively across different versions of the Diffusion Model.

\noindent\textbf{Textual Inversion and DreamBooth with LoRA.}
Textual Inversion teaches new concepts by learning a new token, rather than fine-tuning the entire model. LoRA, on the other hand, employs low-rank weight updates to enhance memory efficiency. The relevant results are shown in Table \ref{tab4}. Our method shows better performance in mitigating the degradation effect on generation with LoRA. Since Textual Inversion does not inject concepts directly into the model, the reduction in quality is not significant and the semantic quality of the generation decrease. 

\noindent\textbf{Main Results}. The results in Table \ref{tab4} demonstrate that adversarial properties of our watermark is preserved across different versions of Stable Diffusion. When the model is mismatched, PIQE exceeds 57.8 and NIQE exceeds 6.37, indicating that our adversarial properties are inheritable across different versions of the diffusion model. Our method ensures that images generated with LoRA have a PIQE higher than 53.82, while the semantic quality metrics of images generated with Textual Inversion, measured by the CLIP score and FID, are 66.47 and 0.292. It effectively resists other types of personalization fine-tuning attacks, such as LoRA and Textual Inversion.

\subsection{Ablation Studies}

\noindent\textbf{Prompt Mismatching.} 
\label{ablation}
Considering that users generate relevant images based on various prompts combined with specific concepts, we generate images for specific concepts under different prompt conditions. The results are shown in Table \ref{fig3}.
% which shows  that our watermarking method achieves higher bit extraction accuracy for images similar to the training set. 
When the image deviates significantly from the set instance prompt, our watermarking method still maintains bit accuacy higher than 0.8. %Furthermore, the generated watermark exhibits a certain level of adversarial robustness, as the quality of images trained under this watermark is reduced, which is beneficial for the protection of specific concepts in diffusion models.

\noindent\textbf{Distortion Layer.} We evaluate the impact of Distortion Layer in the watermark components pretraining. The main results are in Table \ref{tab5}, which indicates that cropping and the mask adversarial layers have a significant impact on image quality, as they require a stronger watermark to be stably embedded within a smaller space. 
% During the pre-training phase, augmentation with JPEG compression and Crop\&Scale significantly enhances the robustness of the watermark. When Crop\&Scale is not included, the watermarking method lacks robustness against such attacks (bit accuracy 0.593). However, after incorporating it, the robustness improves significantly (0.959).

\noindent\textbf{Inference Steps.} 
% Users may influence watermark extraction under different inference parameters. 
% We generate images of specific concepts at different inference steps. 
As shown in Table \ref{tab3}, the bit accuracy remains stable at approximately 0.93 across different settings - specifically reaching 0.917 at 25 steps, 0.936 at 50 steps, and 0.936 at 100 steps. The results indicate that the ConceptWM remains effective  with varying inference steps.

\noindent\textbf{Guidance Scales.}
Our experiments demonstrate that ConceptWM maintains robust watermark detection performance across varying guidance scales, achieving 0.932 bit accuracy at scale 5, 0.917 at scale 7.5, and 0.933 at scale 10.
% Larger guidance scales result in more faithful of the generated image adherence to prompts. We evaluate the impace of guidance scale in the Table \ref{tab3}. The results show that our method maintains stable across various values of the Guidance scale. 

\noindent\textbf{Sampling Method.}
% We evaluate our watermark on DDIM \cite{song2020denoising}, DPM-Solver multi-step \cite{lu2022dpm}, Euler and Heun samplers \cite{karras2022elucidating}. The results show that our watermark is minimally affected by the choice of sampler. 
Evaluation on DDIM \cite{song2020denoising}, DPM-Solver \cite{lu2022dpm}, Euler and Heun samplers \cite{karras2022elucidating} shows consistent performance with 0.929, 0.938, 0.922 and 0.947 bit accuracy respectively, demonstrating minimal impact from sampler choice.

% The watermark detection performs best when using the DPM-Solver multi-step \cite{lu2022dpm} sampler.

\noindent\textbf{Different VAE Decoder.}
% An adversary may interfere with watermark decoding by replacing the VAE decoder. 
%To evaluate the impact of the generated images on the watermark, 
% We used different types of VAEs including stable diffusion V1.4 and V2.1 for decoding. The results in Table \ref{tab3} indicate that changing the VAE does not have a significant impact on the watermark extraction, which is a clear advantage over VAE-based schemes, as the watermark cannot be preserved when the VAE is replaced in VAE-based methods.
An adversary may interfere with watermark decoding by replacing the VAE decoder. We used different types of VAEs including stable diffusion V1.4 and V2.1 for decoding, achieving 0.901 and 0.939 bit accuracy respectively. The results in Table \ref{tab3} indicate that changing the VAE does not have a significant impact on the watermark extraction, which is a clear advantage over VAE-based schemes, as the watermark cannot be preserved when the VAE is replaced in VAE-based methods.

\noindent\textbf{Generation Size.}
% The results in Table \ref{tab3} show that different generation sizes have a noticeable impact on watermark extraction. While the performance at a generation size of 512x512 (bit accuracy 0.933) is significantly better than at 768x768 (bit accuracy 0.812), 
The results in Table \ref{tab3} show that different generation sizes have some impact on watermark accuracy, with 0.933 for 512×512 and 0.812 for 768×768 generations, while the watermarking method still remains effective and detectable across different generation sizes.

%% file: tab/table1.tex
\begin{table*}[]
\begin{tabular}{cccccccccc}
\hline
\multirow{2}{*}{Type} &
  \multirow{2}{*}{Method} &
  \multirow{2}{*}{\begin{tabular}[c]{@{}c@{}}Concept\\ Tracing\end{tabular}} &
  \multirow{2}{*}{\begin{tabular}[c]{@{}c@{}}Concept\\ Protection\end{tabular}} &
  \multicolumn{2}{c}{Fidelity} &
  \multicolumn{4}{c}{Robustness} \\ \cline{5-10} 
 &
   &
   &
   &
  FID &
  CLIP &
  BitAcc  &
  BitAcc(Adv.)  &
  TPR  &
  TPR (Adv.)\\ \hline
 &
  None &
  - &
  - &
  24.37 &
  0.367 &
  - &
  - &
  - &
  - \\ \hline
\multirow{2}{*}{Post-Processing} &
  DwtDctSvd &
  $\times$ &
  $\times$ &
  25.01 &
  0.363 &
  99.98 &
  71.47 &
  1.000 &
  0.367 \\
 &
  RivaGan &
  $\times$ &
  $\times$ &
  24.72 &
  0.363 &
  98.47 &
  83.21 &
  0.994 &
  0.651 \\ \hline
Latent-based &
  Tree-Ring &
  $\times$ &
  $\times$ &
  25.69 &
  0.364 &
  - &
  - &
  1.000 &
  0.821 \\ \hline
VAE-based &
  StableSig. &
  $\times$ &
  $\times$ &
  24.92 &
  0.362 &
  98.49 &
  81.94 &
  0.993 &
  0.848 \\ \hline
\multirow{2}{*}{Concept-Oriented} &
  OLW-based &
  \checkmark &
  \checkmark &
  24.06 &
  0.357 &
  98.68 &
  90.03 &
  0.998 &
  0.884 \\
 &
  FLW-based &
  \checkmark &
  \checkmark &
  24.58 &
  0.365 &
  95.68 &
  89.32 &
  0.994 &
  0.858 \\ \hline
\end{tabular}
\caption{A comparison between our method and previous watermarking techniques is presented. 
%The capacities for DWT-DCT-SVD, RivaGAN, StableSignature, and our method are 48 bits, 32 bits, 48 bits, and 48 bits, respectively. 
We control the false positive rate (FPR) at $10^{-5}$ and evaluate the true positive rate (TPR). Since Tree-ring is a zero-bit watermark, bit accuracy cannot be calculated in this case. ``Adv." (Adversarial) refers to the average performance of images under various distortions.}
\label{tab1}
\end{table*}

%% file: tab/table2.tex
\begin{table*}[t]
\begin{tabular}{ccccccccccc}
\hline
\multirow{2}{*}{Metrics} & \multirow{2}{*}{Methods} & \multicolumn{8}{c}{Distortions}                                                   &         \\ \cline{3-11} 
                         &                          & Contrast & Brightness & Blur  & Noise & Jpeg  & C.\&R. & Vae.C & Denoising & Average \\ \hline
\multirow{4}{*}{Bit Accuracy(\%)}   & Dwtdctsvd & 88.41 & 86.80 & 92.42 & 65.87 & 83.32 & 54.02 & 55.74 & 49.53 & 72.01 \\
                         & RivaGan                  & 97.96    & 94.67      & 94.89 & 91.09 & 90.68 & 94.17         & 70.62 & 58.64     & 86.59   \\
                         & Stablesig.               & 97.86    & 95.48      & 89.88 & 78.94 & 84.40 & 97.47         & 69.94 & 52.58     & 83.31   \\
                         & Ours       & 94.26    & 93.46      & 93.68 & 92.42 & 93.63 & 90.26         & 90.12 & 84.63     & 91.34   \\ \hline
\multirow{5}{*}{Tpr(Fpr=$10^{-5}$)} & Dwtdctsvd & 0.894 & 0.752 & 0.943 & 0.054 & 0.734 & 0.008 & 0.010 & 0.000 & 0.424 \\
                         & RivaGan                  & 0.989    & 0.956      & 0.979 & 0.926 & 0.928 & 0.956         & 0.284 & 0.010     & 0.754   \\
                         & Stablesig.               & 0.987    & 0.985      & 0.905 & 0.369 & 0.821 & 0.989         & 0.293 & 0.000     & 0.669   \\
                         & Tree-ring                & 1.000    & 1.000      & 0.954 & 0.701 & 0.905 & 0.121         & 0.958 & 1.000     & 0.830   \\
                         & Ours        & 0.968    & 0.951      & 0.947 & 0.958 & 0.949 & 0.912         & 0.910 & 0.734     & 0.916   \\ \hline
\end{tabular}
\caption{The performance of ConceptWM against various types of attacks is compared with other watermarking schemes. The attacks include (1) Brightness, (2) Contrast, (3) Blur, (4) Gaussian Noise, (5) JPEG Compression, (6) Crop \& Scale, (7) VAE Compression \cite{cheng2020learned} and (8) Diffusion-based Reconstructive Attack \cite{zhao2023invisible}.}
\label{tab2}
\end{table*}

%% file: tab/table3.tex
\begin{table}[]
\centering
\begin{tabular}{llcc}

\hline
\multicolumn{2}{c}{Settings}             & Bit Accuracy & TPR   \\ \hline
\multirow{3}{*}{Steps}           & 25    & 0.917  & 0.921 \\
                                 & 50    & 0.936  & 0.946 \\
                                 & 100   & 0.936  & 0.951 \\ \hline
\multirow{4}{*}{Sampler}         & DDIM  & 0.929  & 0.939 \\
                                 & DPM-M & 0.938  & 0.972 \\
                                 & EULER & 0.922  & 0.941 \\
                                 & HEUN  & 0.947  & 0.962 \\ \hline
\multirow{3}{*}{Guidance Scale}  & 5     & 0.932  & 0.948 \\
                                 & 7.5   & 0.917  & 0.932 \\
                                 & 10    & 0.933  & 0.947 \\ \hline
\multirow{2}{*}{Generation Size} & 512   & 0.933  & 0.951 \\
                                 & 768   & 0.812  & 0.748 \\ \hline
\multirow{2}{*}{VAE}             & V2.1  & 0.939  & 0.954 \\
                                 & V1.4  & 0.901  & 0.913 \\ \hline
\end{tabular}
\caption{The impact of inference settings on ConceptWM watermark detection.}
% Our concept watermarking approach remains effective across various generation steps, different samplers, VAE versions, and generated shapes.}
\label{tab3}
\end{table}

%% file: tab/table4.tex
\begin{table*}[]
\centering
\begin{tabular}{cccccclcccl}
\hline
\multirow{2}{*}{} &
  \multirow{2}{*}{TRAIN} &
  \multirow{2}{*}{TEST} &
  \multicolumn{4}{c}{``a photo of $\langle$concept$\rangle$ person"} &
  \multicolumn{4}{c}{``a portrait of $\langle$concept$\rangle$ person"} \\ \cline{4-11} 
                                                                          &      &      & PIQE↓ & NIQE↓ & CLIP↑ & FID↓  & PIQE↓ & NIQE↓ & CLIP↑ & FID↓  \\ \hline
\multirow{2}{*}{\begin{tabular}[c]{@{}c@{}}MODEL\\ MISMATCH\end{tabular}} & v2.1 & v1.4 & 63.18 & 6.87  & 0.287 & 87.21 & 54.7  & 6.79  & 0.272 & 84.53 \\
                                                                          & v1.4 & v2.1 & 59.82 & 6.72  & 0.293 & 72.18 & 57.8  & 6.37  & 0.292 & 66.28 \\ \hline
LORA                                                                      & v2.1 & v2.1 & 53.82 & 5.72  & 0.306 & 69.48 & 55.82 & 5.37  & 0.311 & 70.98 \\
Textual Inversion                                                         & v2.1 & v2.1 & 36.87 & 4.96  & 0.294 & 66.47 & 39.58 & 4.48  & 0.285 & 56.83 \\ \hline
\end{tabular}
\caption{An experiment on generating images with adversarial watermarks and fine-tuning them with different model versions and methods for personalization. The effectiveness of our adversarial watermark defense through personalized fine-tuning is observed across different model versions and personalized methods.}
\label{tab4}
\end{table*}

%% file: tab/table5.tex
\begin{table}[]
\begin{tabular}{ccccc}
\hline
 & PSNR & SSIM & \begin{tabular}[c]{@{}c@{}}BitAcc\\ (w/)\end{tabular} & \begin{tabular}[c]{@{}c@{}}BitAcc\\ (w/o)\end{tabular} \\ \hline
GaussianNoise & 33.26 & 0.928 & 0.962 & 0.934 \\
ColorJitter   & 33.78 & 0.939 & 0.979 & 0.972 \\
GaussianBlur  & 33.69 & 0.931 & 0.967 & 0.948 \\
Jpeg          & 32.26 & 0.933 & 0.969 & 0.882 \\
CROP\&SCALE   & 30.58 & 0.917 & 0.959 & 0.593 \\
Random Mask   & 32.19 & 0.915 & 0.969 & 0.908 \\
Combination   & 29.08 & 0.889 & 0.954 & 0.528 \\ \hline
\end{tabular}
\caption{The impact of various attacks on image quality in the distortion layer during pretraining.}
\label{tab5}
\end{table}

%% file: sec/6_conclusion.tex
% \section{Discussion}
% \textbf{Limitations.} First, our method exhibits limited detection performance under certain prompts. As shown in Figure \ref{fig3}, when the model generates poor results for specific concepts under certain prompts, the detection accuracy cannot be guaranteed. This issue requires more precise and improved personality techniques or larger datasets and prompts. Second, the adversarial perturbations generated by our proposed method have limited effectiveness against the latest purification methods \cite{zhao2024can,299878}. Since our approach involves embedding concepts, watermarks, and adversarial perturbations into the diffusion model, there are more constraints to consider, which makes our adversarial perturbations more limited compared to the state-of-the-art image adversarial perturbation methods \cite{liang2023adversarial,van2023anti}. In the future, we will explore ways to better embed adversarial perturbations into the Diffusion model.

\section{Conclusion}
This paper reveals the potential threats posed by diffusion model personalization techniques and proposes an adversarial watermarking method to protect specific concepts. 
Our solution integrates adversarial perturbations and watermarks into the diffusion model, ensuring that images generated by the diffusion model for specific concepts are traceable, and any personalized model trained on these images will produce low-quality results. 
% We have evaluated the stability and adversarial effectiveness of the watermarking method, demonstrating that our defense remains effective under adverse conditions. 
% In the future, our main focus will be on two aspects of improvement. On the one hand, 
In the future, The investigation of concept-specific learning can be extended beyond DreamBooth, necessitating an in-depth examination of watermark embedding paradigms for textual inversion, LoRA, and related concept learning methodologies. 
% we aim to address the efficiency issue caused by the need to generate adversarial noise in the current method. On the other hand, we will consider potential threats from attackers who may target the model level.